\documentclass{ws-procs961x669}            
\usepackage{siunitx}
\usepackage[dvipsnames]{xcolor}

\begin{document}
\title{The COSmic Monopole Observer (COSMO)}

\author{S. Masi$^*$, E. Battistelli, P. de Bernardis, A. Coppolecchia, F. Columbro, G. D'Alessandro, \\ M. De Petris, L. Lamagna, E. Marchitelli, L. Mele, A. Paiella, F. Piacentini, G. Pisano}

\address{Physics Department, Sapienza University of Rome\\
and INFN Sezione di Roma, P.le A. Moro 2, Rome, 00185, Italy\\
$^*$e-mail: silvia.masi@roma1.infn.it\\
https://cosmo.roma1.infn.it}

\author{M. Bersanelli, C. Franceschet, E. Manzan, D. Mennella, S. Realini}

\address{Physics Department, University of Milan\\
and INFN Sezione di Milano, Via Celoria 16, 20133  Milan, Italy\\}

\author{S. Cibella, F. Martini, G. Pettinari}

\address{IFN-CNR\\
Via Cineto Romano 42, 00156 Roma, Italy\\}

\author{G. Coppi, M. Gervasi, A. Limonta, M. Zannoni}

\address{Physics Department, University of Milan Bicocca\\
and INFN Sezione di Milano Bicocca, P.zza delle Scienze 1, 20126 Milan, Italy\\}

\author{L. Piccirillo}

\address{Department of Physics and Astronomy, University of Manchester\\
Oxford road, M13 9PL Manchester, UK\\}

\author{C. Tucker}

\address{School of Physics and Astronomy, University of Cardiff\\
The Parade, CF24 3AA Cardiff, UK\\}

\begin{abstract}
The COSmic Monopole Observer (COSMO) is an experiment to measure low-level spectral distortions in the isotropic component of the Cosmic Microwave Background (CMB). Deviations from a pure blackbody spectrum are expected at low level ($<$ 1 ppm) due to several astrophysical and cosmological phenomena, and promise to provide important independent information on the early and late phases of the universe. They have not been detected yet, due to the extreme accuracy required, the best upper limits being still those from the COBE-FIRAS mission. COSMO is based on a cryogenic differential Fourier Transform Spectrometer, measuring the spectral brightness difference between the sky and an accurate cryogenic blackbody. The first implementation of COSMO, funded by the Italian PRIN and PNRA programs, will operate from the Concordia station at Dome-C, in Antarctica, and will take advantage of a fast sky-dip technique to get rid of atmospheric emission and its fluctuations, separating them from the monopole component of the sky brightness. Here we describe the instrument design, its capabilities, the current status. We also discuss its subsequent implementation in a balloon-flight, which has been studied within the COSMOS program of the Italian Space Agency.
\end{abstract}


\bodymatter

\section{Introduction}\label{intro}

\begin{figure}[b]
\centering
\includegraphics[angle=180, width=\textwidth]{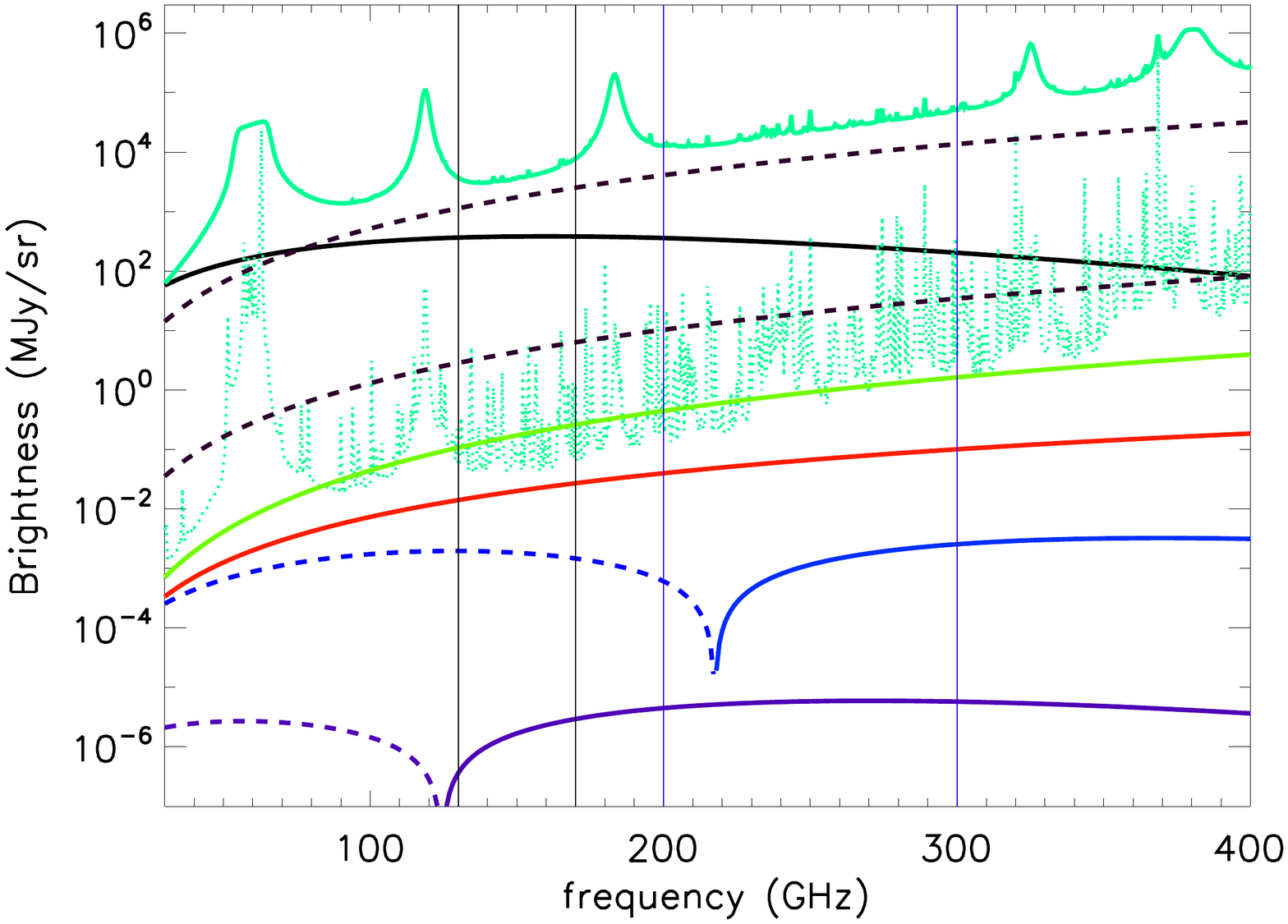}
\caption{Plot of isotropic brightness from (top to bottom, continuous lines): Antarctic atmosphere in Dome-C (zenith), CMB, interstellar dust (Galactic average), far infrared background (FIRB) from unresolved galaxies, absolute value of the $y$-distortion ($y=1.77\times10^{-6}$, negative branch dashed), absolute value of the $\mu$-distortion ($\mu=2\times 10^{-8}$, $\beta=2.19$,  negative branch dashed). The two dashed lines represent the emission of the window of the receiver, for the ground based (upper) and the balloon-borne (lower) implementations of COSMO. The dotted line represents atmospheric emission at stratospheric balloon altitude (zenith). The two couples of vertical lines define the observation bands of COSMO in its ground-based implementation.} 
\label{fig:spectra}
\end{figure}

Spectral distortions of the CMB represent a research path orthogonal and synergic to CMB polarization studies. 
Their detection can shed light on cosmic reionization, on the physics of recombination, on dark matter and in general on any energy release in the primeval fireball, and even on the very early universe and cosmic inflation. See \cite{Chluba21} and references therein for a comprehensive review.  The expected deviations are very small, $<$ 1 ppm, but have characteristic spectral signatures, allowing, in principle, their separation from overwhelming foreground emission and the CMB monopole itself. In figure \ref{fig:spectra} we compare the largest spectral distortions of the CMB in the mm-wave region to the most important foregrounds. 

Current upper limits for spectral distortions are at a level of 0.01\% of the peak brightness of the CMB, and were obtained more than 20 years ago by COBE – FIRAS \cite{Mather90, Fixsen96}. Current technologies promise to achieve much higher sensitivity, the limiting factor remaining the control of instrumental systematic effects and the emission of astrophysical foregrounds. In order to avoid the very large background produced by the earth atmosphere at CMB frequencies, the final measurement must be carried out from space. To this purpose, several proposals for space-based experiments have been submitted, like the PIXIE \cite{Kogut11}, CORE \cite{debe18} and PRISTINE missions. Meanwhile, ground-based and near-space efforts are necessary to test and refine methods, and possibly to detect the largest spectral distortions. Here we focus on the COSMO project, a pathfinder, staged effort to be carried out in Dome-C (Antarctica) first, and then on a stratospheric balloon. 

The problem of spectral distortions of the CMB monopole, as evident from figure \ref{fig:spectra}, is that they are very small with respect to other isotropic backgrounds, including instrument emission, atmospheric emission (if present), astrophysical foregrounds from the interstellar medium and the other galaxies along the line of sight, and the CMB monopole itself.  Moreover the instrument has to be cryogenic to reduce its own emission, and operate where atmospheric emission is very small and/or can be measured and subtracted. For sure, advanced strategies for components separation are in order (see e.g. \cite{Chluba17}).

In the following section we focus on the COSMO experiment, attempting sub-orbital measurements of spectral distortions.

\section{The COSMO experiment} 

\subsection{Measurement method and general design}

COSMO is a cryogenic differential Fourier Transform Spectrometer (FTS) in Martin Puplett Interferometer (MPI) configuration \cite{Mart70}: a two input ports differential instrument comparing the brightness of the sky (port A) to the brightness of an accurate internal blackbody (port B), exactly like in COBE-FIRAS\cite{Fixsen94}. A simplified instrument model is described by equation \ref{eq:instrument}:
\begin{equation}\label{eq:instrument}
I_{sky}(x) = {\cal R} \int_0^\infty A\Omega (\sigma) [ B_{sky}(\sigma) - B_{ref}(T_{ref}, \sigma) ] e(\sigma) [ 1 + cos(4 \pi \sigma x) ] d\sigma
\end{equation}
where $I_{sky}(x)$ is the signal measured by the detector as a function of the optical path difference $x$ between the two arms of the interferometer, which is scanned by means of a moving roof-mirror;  $\sigma$ is the wavenumber, $A\Omega (\sigma)$ is the optical throughput of the detector, $B_{sky}(\sigma)$ is the sky brightness present at input port A, $B_{ref}(T_{ref}, \sigma)$ is the brightness of the reference blackbody at temperature $T_{ref}$ present at input port B, $e(\sigma)$ is the spectral efficiency of the instrument, the constant $\cal R$ is the responsivity of the instrument. 
During the calibration phase, an external blackbody at a known temperature $T_{ext}$ fills input port A: in these conditions the measured signal is 
\begin{equation}\label{eq:calibration}
I_{cal}(x) = {\cal R} \int_0^\infty A\Omega (\sigma) [ B_{ext}(T_{ext}, \sigma) - B_{ref}(T_{ref}, \sigma) ] e(\sigma) [ 1 + cos(4 \pi \sigma x) ] d\sigma .
\end{equation}
Antitransforming equation \ref{eq:calibration} one gets the quantity ${\cal R} A\Omega (\sigma) e(\sigma)$, which can then be used in the antitransform of equation \ref{eq:instrument} to estimate $B_{sky}(\sigma)$. 

Both $B_{ref}$ and $B_{ext}$ should be at cryogenic temperatures, close to $T_{CMB}$, in order to avoid saturation or non-linearity effects in the detectors, and extreme requirements on the knowledge of $T_{ref}$, $T_{ext}$ and the blackbody emissivities.

\begin{figure}[b]
\includegraphics[width=\textwidth]{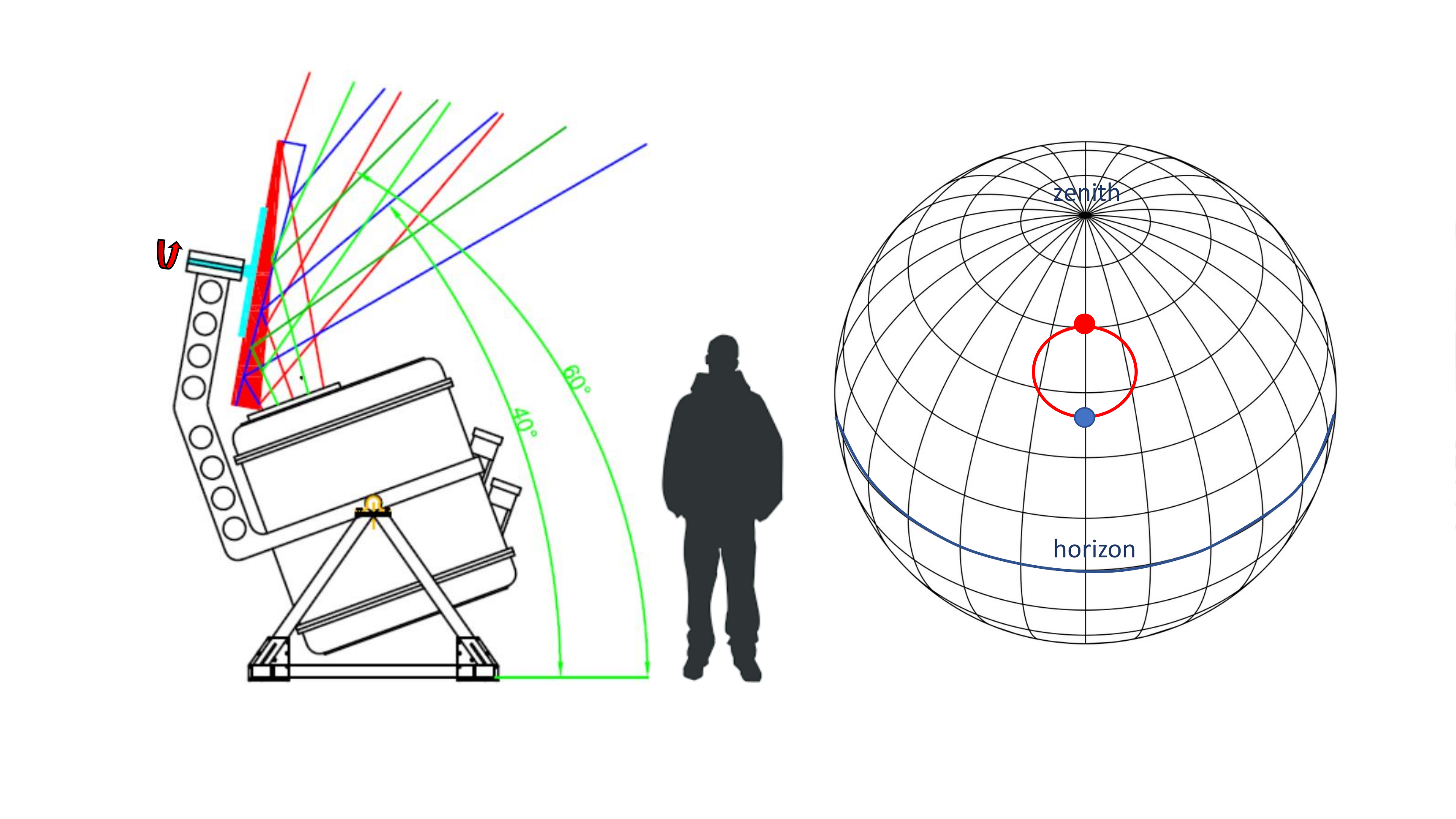}
\caption{Geometry of the sky scan for the COSMO experiment. {\bf Left:} The COSMO instrument with the external spinning wedged flat mirror, steering the instrument beam in the sky while spinning. The arrow identifies the spin axis, while the mirror is sketched in two positions (red and blue) corresponding to maximum and minimum elevation of the beams, respectively. The central and marginal beams are also shown in the same colors. {\bf Right:} Resulting scan of the beam (in red) over the celestial sphere. The two spots mark the maximum and minimum elevations explored by the beam. The circular scans, combined with Earth rotation, result in the coverage of several sky patches at different Galactic latitudes.}
\label{fig:scan}
\end{figure}

Since the COSMO instrument operates in environments (Dome-C or the stratosphere) where atmospheric emission is present, special care must be taken to measure the emission from the atmosphere, and the emission of the vacuum window separating the cryogenic instrument from the external environment. Both these emissions are not present in a space-based instrument, while they contribute very significantly to the measured $B_{sky}$ in the case of a suborbital experiment like COSMO. Moreover, the cryogenic external blackbody will necessarily have a vacuum window, adding its emission to $B_{ext}$.

\subsection{Coping with atmospheric emission}\label{atmosphere}

In the ground-based implementation, COSMO will operate from the French-Italian Concordia base, in Dome-C (Antarctica). This is likely to be one of the best sites on Earth for mm-wave astronomical observations, being extremely cold and dry. But still atmospheric emission is present. The first defense against atmospheric emission is the selection of the frequency interval. In its ground-based implementation, COSMO uses two frequency channels matching mm-wave atmospheric windows, covering the frequency bands 110-170 GHz and 200-300 GHz. Each frequency channel is detected by an independent focal plane array. In this way radiation from the high frequency band does not overload the low-frequency detectors, so that higher sensitivity can be achieved in the low frequency channel.

COSMO uses fast Kinetic Inductance Detectors (KID) and fast elevation scans to separate atmospheric emission and its long-term fluctuations from the monopole of the sky brightness. 
A fast spinning, wedged flat mirror ($>$1000 rpm) steers the boresight direction on a circle, 20$^o$ in diameter, scanning a range of elevations (and the corresponding optical depths of the atmosphere) while the cryogenic interferometer scans the optical path difference (see figure \ref{fig:scan}).   

\begin{figure}[b]
\includegraphics[width=\textwidth]{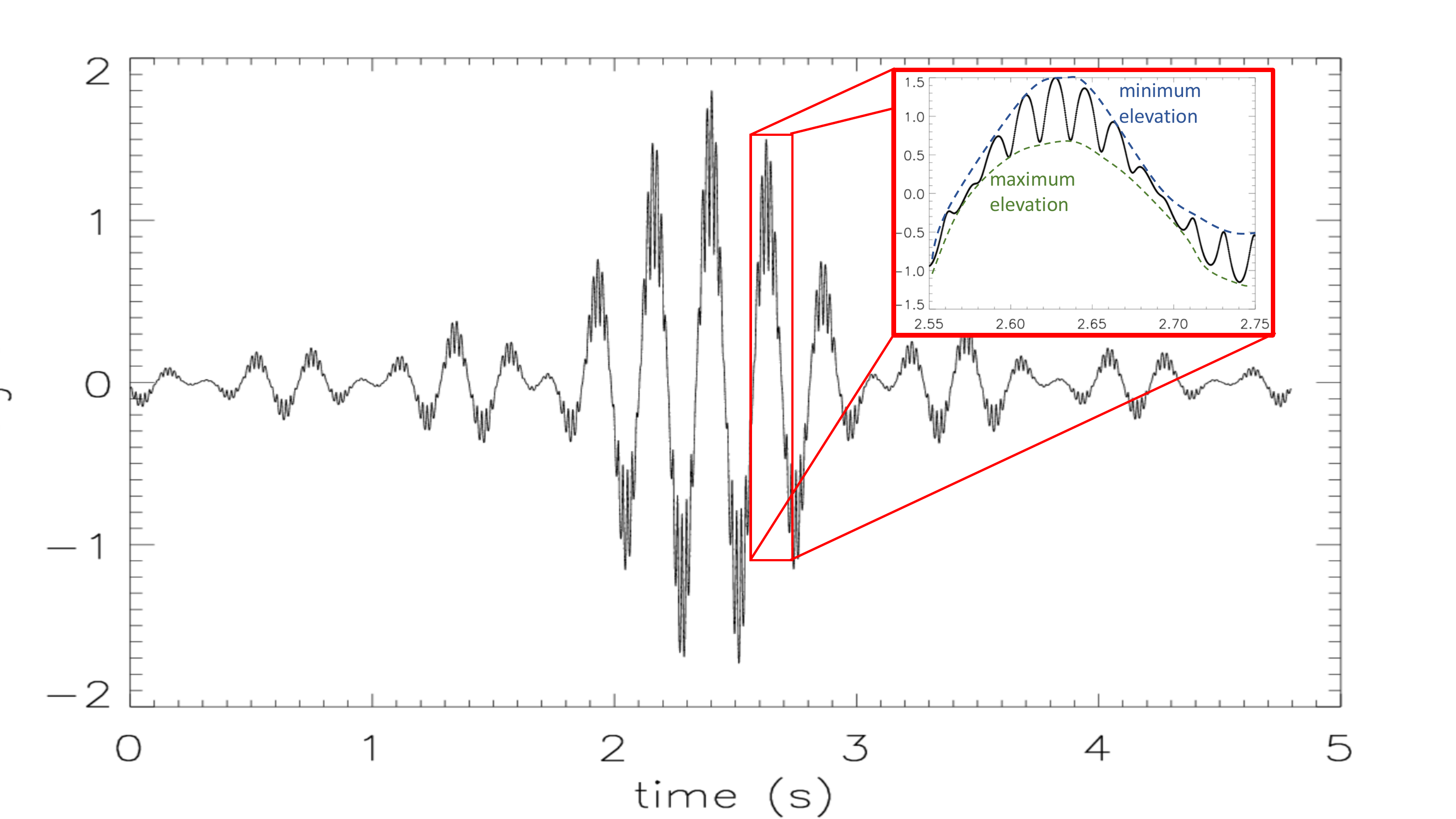}
\caption{Example of expected interferogram obtained while the flat wedged mirror is spinning. For each position $x$ of the roof mirror in the MPI (corresponding to a different time on the abscissa), the signal changes due to the fast variation of the observed elevation. Fast (KID) detectors are used to follow these variations. In this example one full scan of the roof mirror is shown (at a mechanical speed of $\sim$ 0.25 cm/s, from -1.27 cm to +1.27 cm of optical path difference). The zoomed inset shows how the interferograms corresponding to the maximum and minimum elevations can be obtained by properly resampling the measured interferogram. Millions of interferograms like this one will be measured during an observation campaign.}
\label{fig:interferogram}
\end{figure}

In this way the recorded interferogram (equation \ref{eq:instrument}) will be the one due to the sky monopole (constant during the sky scan) plus the one due to atmospheric emission, which is modulated by the beam elevation variations during the sky scan. A sample expected interferogram is reported in figure \ref{fig:interferogram} for the low-frequency band of COSMO. 

\begin{figure}[b]
\includegraphics[width=\textwidth]{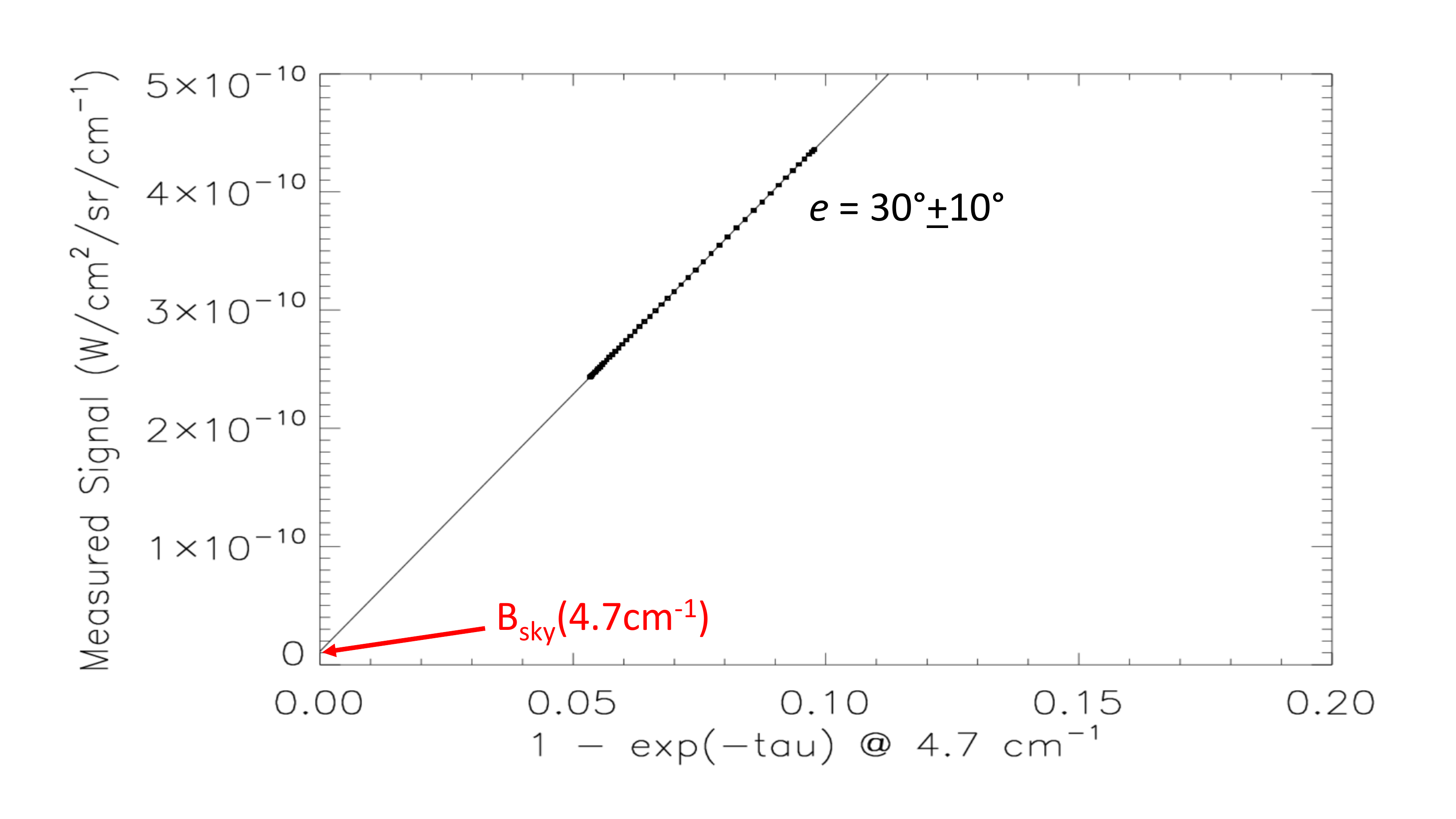}
\caption{Extrapolation to zero air-mass to estimate the isotropic component of the sky brightness for each frequency.}
\label{fig:cosec}
\end{figure}

By properly sampling the measured interferograms, the interferograms corresponding to the different elevations $e$ scanned by the instrument can be retrieved. These can be antitrasformed to estimate the measured brightness spectra for the different elevations. For each frequency, a cosec($e$) law can be used to estrapolate to zero air-mass, thus estimating the isotropic component of the sky brightness (see figure \ref{fig:cosec}). 

Since the instrument measures one interferogram every few seconds, slower atmospheric fluctuations are removed in the process, mitigating the effects of $1/f$ noise, which is characteristic of atmospheric fluctuations. 

This measurement method is allowed by the use of fast ($\tau \sim 60 \mu s$) KIDs, similar to the ones developed for the OLIMPO experiment\cite{Masi19}. In table \ref{tab:instrument} we report a possible scanning configuration of the COSMO instrument in its first ground-based implementation. 

\begin{table}
\tbl{Scanning characteristics of the COSMO instrument$^{\text a}$}
{\begin{tabular}{@{}lrr@{}}
\toprule
circle radius \hphantom{00} & \hphantom{0}5 & \hphantom{0} deg \\
beam FWHM \hphantom{00} & \hphantom{0}0.5 & \hphantom{0}deg  \\
wedged mirror spin \hphantom{00} & \hphantom{0} 600 & \hphantom{0} rpm \\
time per beam \hphantom{0} & \hphantom{0}200 & \hphantom{0}$\mu$s \\
time for one forward plus one reverse sky dip & 0.1 & \hphantom{0}s\\
\colrule
maximum wavenumber \hphantom{00} & \hphantom{0}20 & \hphantom{0} cm$^{-1}$ \\
sampling step \hphantom{00} & \hphantom{0}125 & \hphantom{0}$\mu$m  \\
resolution \hphantom{00} & \hphantom{0} 5-15 & \hphantom{0} GHz \\
time to complete one interferogram \hphantom{0} & \hphantom{0}25.6 & \hphantom{0}s \\
sky dips per interferogram element \hphantom{0} & \hphantom{0}2 & \hphantom{0} \\
\botrule
\end{tabular}
}
\begin{tabnote}
$^{\text a}$ Example of certainly feasible combination of wedge and roof mirror scans; faster scans are currently under consideration.\\
\end{tabnote}
\label{tab:instrument}
\end{table}

\subsection{Coping with window emission}\label{window}

Both the cryogenic reference blackbody and the FTS of COSMO are enclosed in a vacuum shell, necessary for the cryogenic operations. External radiation enters the vacuum jacket through a vacuum window, made from a transparent material for mm waves. The best solution is to use a slab of polypropylene, which has to be $\sim$ 1 cm thick in the ground based implementation, to withstand atmospheric pressure. This can be significantly thinner at balloon altitude: we assume a 25 $\mu$m thick window for that implementation. As visible in figure \ref{fig:spectra}, the emission of the receiver window is very important in both cases. The way we cope with it is to measure spectra keeping the internal reference and the external calibrator at the two input ports and steady, for different, well controlled temperatures of the window. This procedure will produce a calibration of the contribution from window emission in the measured spectra as a function of the temperature of the window. This will be monitored during the sky measurements, so that will be possible to subtract window emission using the calibration. The expected accuracy of this process has been investigated: preliminary results show that the residuals after subtraction are subdominant when targeting at the measurement of the $y$ distortion.

\section{Implementation of the ground-based COSMO experiment}\label{groundimplementation}

\subsection{Selection of the site and installation}

Due to the high altitude, extremely low temperatures in the winter, and absence of solar illumination for several months in a row, the Dome-C (Antarctica) site offers unmatched excellent conditions for astronomical observations in the mm-wave atmospheric windows (see e.g. \cite{Tremblin11, Battistelli12}). Its latitude (75$^o$S) allows for crosslinked sky scans at constant elevation (which are not possible from the South Pole), similar to those obtained in circumpolar stratospheric balloon flights (see e.g. \cite{Masi06}).  The presence of the French-Italian Concordia station provides high quality logistics support (power, communications, resident personnel) for the operations of a modern CMB experiment. The instrument will be hosted in a thermally insulated ISO20 shipping container, with retractable roof, and mounted on a palafitte on top of a hardened snow berm, following a consolidated tradition for medium-size astronomical instruments in Antarctica. The container and the instrument will be oriented so that the elevation movement steers the boresight along the local meridian. As described below (see \S \ref{sec:performance}), this simple sky scan strategy allows to observe both high and intermediate Galactic latitudes, totalizing a coverage of $\sim 5 \%$ of the sky. 

\subsection{The optical system}

\subsubsection{Generalities}

\begin{figure}
\includegraphics[width=\textwidth]{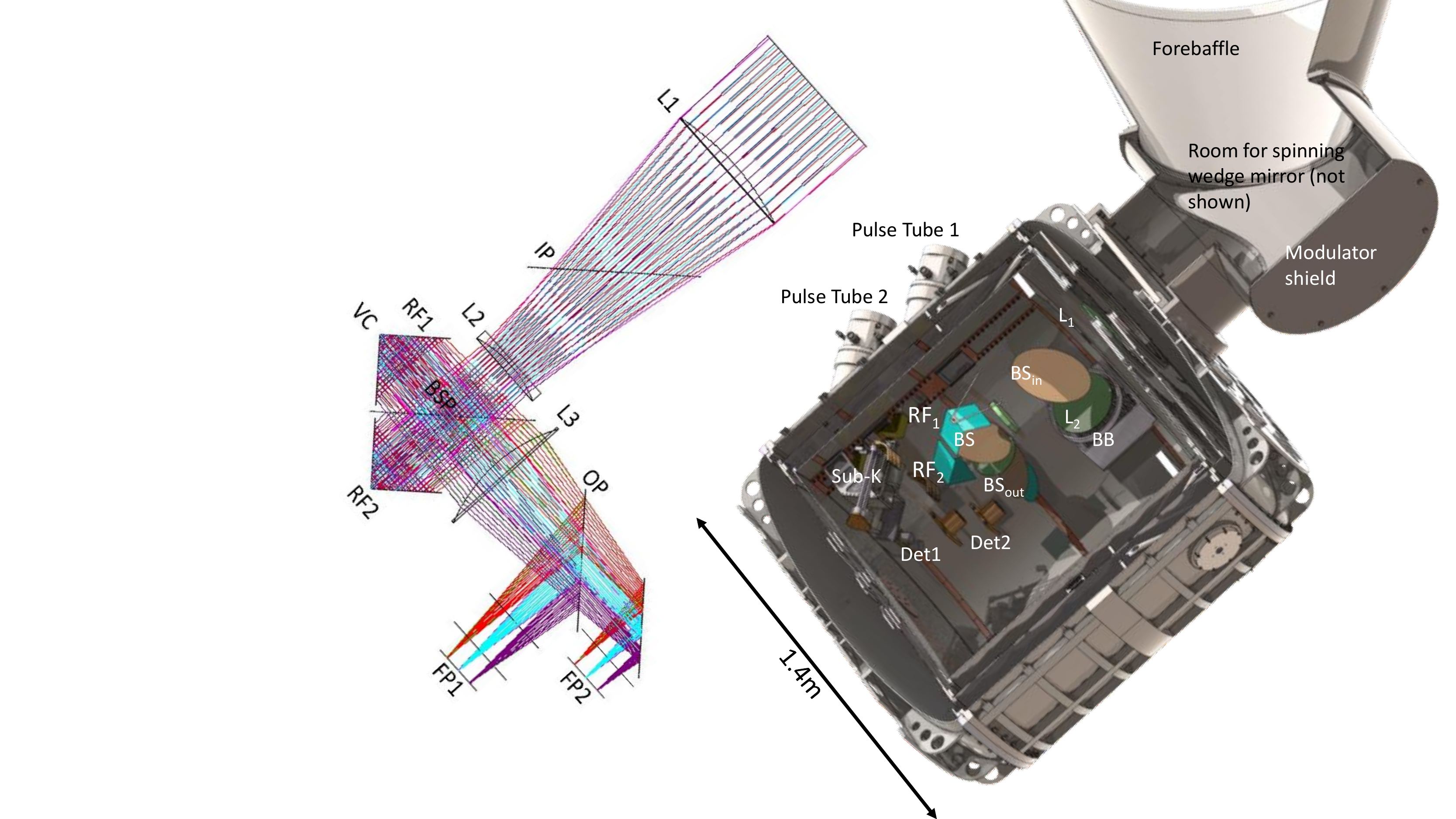}
\caption{{\bf Left:} Ray tracing and optical design of COSMO. L1 is the light collector lens, L2 produces a nearly parallel beam for the FTS, L3 focuses radiation of the focal planes FP1 and FP2. The Differential Fourier Transform Spectrometer is costituted by the input polarizer IP, the beam-splitter polarizer BSP, the roof mirrors RF1 and RF2, the output polarizer OP. The input polarizer transmits radiation from the sky and reflects radiation from the cryogenic reference blackbody. {\bf Right:} rendering of the accommodation of the COSMO instrument inside its cryogenic system.}
\label{fig:optics}
\end{figure}

Angular resolution is not the main driver of this experiment, which is aimed at measurements of the {\em isotropic}  monopole component of the CMB brightness. We designed the instrument with and optical aperture of 22 cm in diameter, producing a diffraction limited beam at the longest wavelength (2.4 mm) of $\sim 0.7^o$ FWHM. This is the result of a tradeoff between the size of the optical aperture and the need of monitoring the effect of {\em anisotropic} foregrounds. The telescope and the Fourier Transform Spectrometer are located inside the cryostat to minimize their radiative background. A sketch of the optical setup, based on polyethylene lenses, is shown in the left panel of figure \ref{fig:optics}, while the main data are reported in table \ref{tab:optics}.

\begin{table}
\tbl{Optical parameters of the COSMO instrument.}
{\begin{tabular}{@{}lrr@{}}
\toprule
optical aperture diameter \hphantom{00} & \hphantom{0}220 & \hphantom{0} mm \\
effective focal length \hphantom{00} & \hphantom{0}726 & \hphantom{0} mm \\
multimode pixel antenna aperture diameter \hphantom{00} & \hphantom{0}20 & \hphantom{0} mm \\
focal planes \hphantom{00} & \hphantom{0}2 &  \\
number of detectors per focal plane \hphantom{00} & \hphantom{0}9 &  \\
projected pixel to pixel distance (x and y) \hphantom{00} & \hphantom{0}0.75 &  \hphantom{0}deg \\
beam FWHM \hphantom{00} & \hphantom{0}0.75 & \hphantom{0}deg  \\
\botrule
\end{tabular}
}
\label{tab:optics}
\end{table}

\subsubsection{The cryogenic Fourier Transform Spectrometer}

\begin{figure}[t]
	\includegraphics[width=0.7\columnwidth]{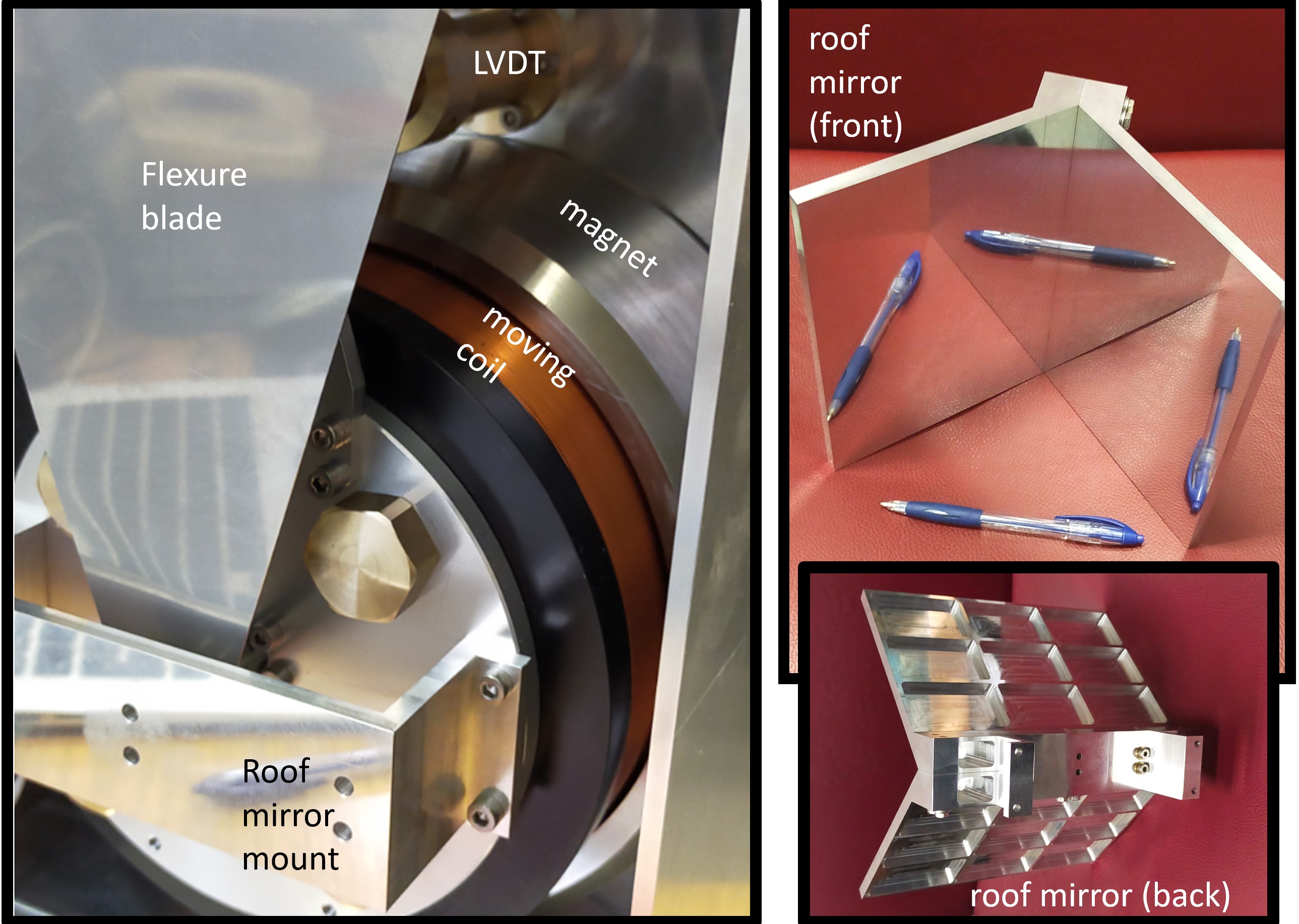}
	\centering
        \caption{{\bf Left:} photo of the cryogenic roof-mirror scanner for the MPI FTS. {\bf Right:} the roof mirror. Its projected size is 200 $\times$ 200 mm.}
     \label{fig:cryomechanism}
    \end{figure}

The cryogenic FTS is a MPI, with two delay lines terminated with roof-mirrors. The interferogram is obtained introducing a variable difference in the lengths of the two delay lines. This is obtained with the linear movement of one of the two roof-mirrors (RF1 in figure \ref{fig:optics}). The maximum optical path difference introduced by the motion of the moving roof mirror is +/- 2 cm, corresponding to a coarse spectral resolution of $\sim$10 GHz. Since the FTS operates at cryogenic temperature, we have to minimize the friction produced by the motion of the roof-mirror. We have designed a cryomechanism based on harmonic steel flexure blades, operating in the elastic regime. The actuator is a coil moving in the radial magnetic field of a strong permanent magnet (figure \ref{fig:cryomechanism}). The position is sensed by a LVDT. Preliminary room-temperature tests confirm the frictionless operation of the device.

\subsubsection{The detectors}

The output beam from the FTS is naturally split in two exit ports by the output polarizer. In each output port we place a focal plane array, consisting of multimode horn antennas feeding multimode lumped elements kinetic inductance detectors (LEKIDs), on a 4 inch Si wafer. The reflected beam serves the high frequency band array, while the transmitted one serves the low frequency band array. 

The design of the COSMO LEKIDs was optimized starting from the experience gained with the LEKID arrays realized for the OLIMPO experiment \cite{Paiella_JCAP, Paiella_electronics}. 
COSMO detectors are \SI{30}{nm} thick Aluminum LEKIDs, each of which has a front--illuminated V order Hilbert curve absorber/inductor. The large absorber area, $\sim\SI{8}{mm}\times\SI{8}{mm}$ for both channels, serves to guarantee the efficient absorption of all the modes which propagate through the multimode waveguides. In particular, the size of the waveguide diameter, \SI{4.5}{mm} for the \SI{150}{GHz} channel and \SI{4}{mm} for the \SI{255}{GHz} channel, allows the propagation of 10 to 19 modes in the 120-180 GHz range and of 23 to 42 modes in the 210-300 GHz range. Because of the size of the focal planes and the horn apertures, each detector array hosts 9 pixels.

The forecast performance for such detectors, in terms of detector noise equivalent power (NEP) is ${\rm NEP_{det}}\sim\SI{3.8e-17}{W/\sqrt{Hz}}$, with a response time $\tau\sim\SI{60}{\micro\second}$, when operated at about \SI{300}{mK}. These estimations take into account the absorber volume, the superconductor critical temperature, and the simulated absorber efficiency. This performance is more than adequate for the ground-based implementation of COSMO, which is likely to be dominated by atmospheric noise and systematic effects rather than by detector sensitivity.

 \begin{figure}[t]
	\includegraphics[width=0.48\columnwidth]{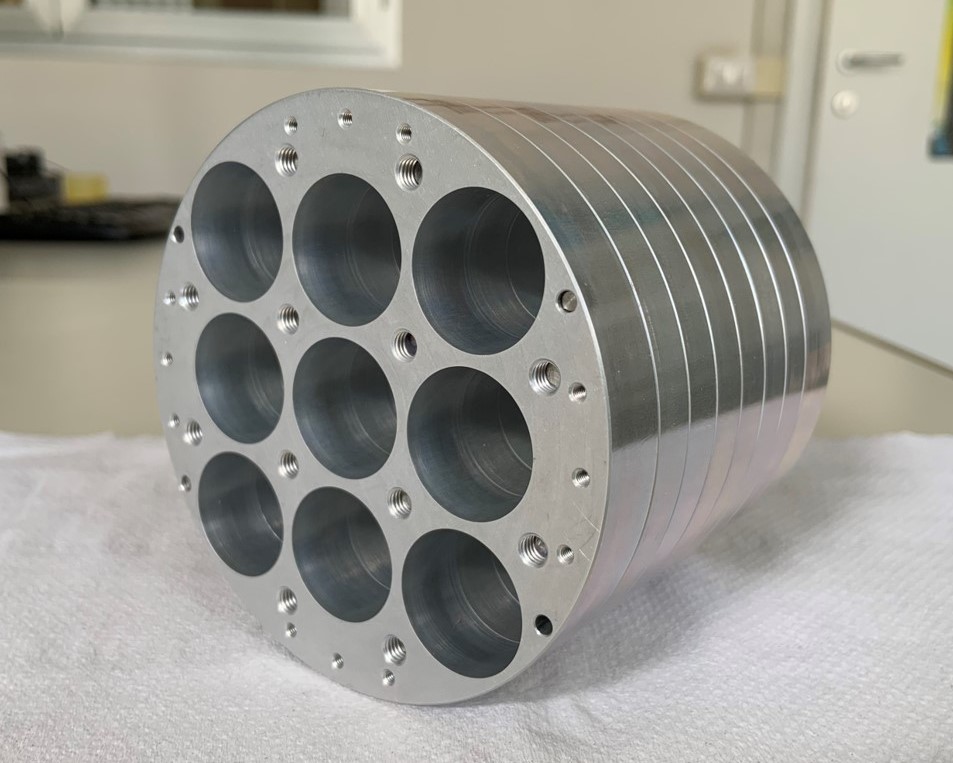}
	\includegraphics[width=0.48\columnwidth]{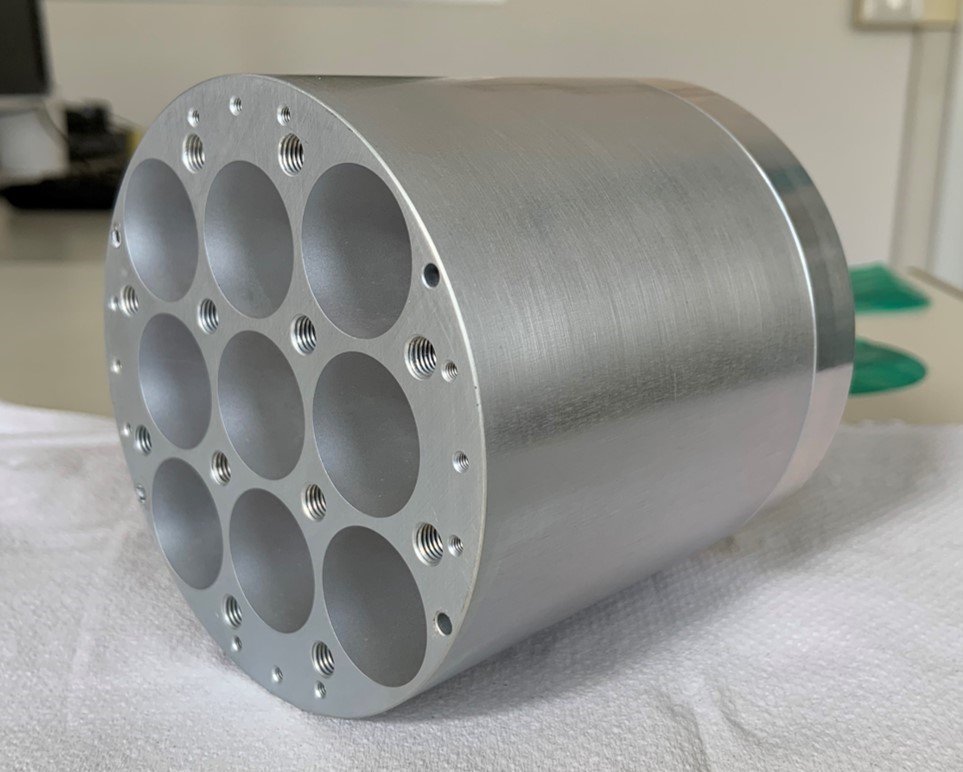}
        \caption{{\bf Left:} photo of the low-frequency array, made of seven antenna plates and a bottom plate with circular waveguides. {\bf Right:} photo of the high-frequency array, which is made as one piece, with a separate bottom plate of circular waveguides.}
     \label{fig:COSMO_Array_images}
    \end{figure}

\subsubsection{The detector feedhorns}

The radiation coming from the FTS is coupled to the detectors by two arrays of nine multimoded feed-horns working in the $120-180$ GHz and $210-300$ GHz range, respectively.
Since COSMO will perform only total intensity measurements, we chose a smooth-walled profile which is easier to manufacture than a corrugated one.


The low-frequency array consists of Winston cone antennas while the high-frequency array hosts linear profiled antennas. The antenna design is the result of a trade-off between the multimode requirement on the antenna waveguide, the mechanical constraint on the antenna aperture and the optimization of the antenna directivity within the cryostat aperture window. The forecasted FWHM of the main beam is around $20^{\circ}$, the first side-lobe level is below -15 dB and the far side-lobe level is below -30 dB.

The arrays are machined with CNC milling by Pasquali SRL in Milan, and are obtained by superimposing metal plates through dowel-pins and tightening them with screws, as shown in figure~\ref{fig:COSMO_Array_images}. The arrays are cylindrical with a diameter of 101 mm to minimize the mass to be cooled as much as possible; the feed-horns are arranged on a square footprint with a 26 mm center-to-center distance.

Both the arrays are made in ergal (Al7075) to conform to the focal plane and avoid differential thermal contractions during the cryogenic cooling of the experiment. They are mechanically coupled to a band-pass filter on the front (aperture) side and to an interface to the focal plane wafer holder on the back. To this purpose, we used the experience gained with the LSPE-SWIPE experiment\cite{Columbro2020SWIPEMP} to design a flared waveguide interface to directly illuminate the KIDs detectors.

\subsubsection{The reference blackbody}

One of the two inputs of FTS looks at a cryogenic reference blackbody. This is a cavity with the inner surface covered by a microwave absorber. The cavity structure is in copper, to achieve a very uniform temperature. We have optimized the shape of the cavity 
both through geometrical ray tracing (maximizing the number of internal reflections) and via finite elements electromagnetic simulations (minimizing the S$_{11}$ parameter). The performance in the ray-tracing approximation is assessed assuming a reflectance of the coating as a function of the wavenumber $\bar{\nu}$ as in ref.\cite{firas_cal} , $r(\bar{\nu})=0.08 + \frac{0.06[cm^{-1}]}{\bar{\nu}}$. The number of internal reflections is always $N>6$, corresponding to the cavity reflectance $R=r^N<10^{-6}$ over the band \SI{100}{}-\SI{300}{\giga \hertz}. This indirectly provides the cavity emissivity as $(1-R)$. 

Due to the size of the calibrator, much larger than the wavelengths of interest, we used \textit{HFSS} electromagnetic analysis only to simulate the performance of a thin cut section of the cavity. This is a few wavelengths thick, and is representative in terms of the absorbing properties of the cavity. However it is not representative of the direction of the damped outgoing radiation. The simulations have been run for the lowest frequencies of the \SI{150}{\giga \hertz} band, and the corresponding dielectric properties of the absorbing coating (relative permittivity $\epsilon_r=3.50$ and dielectric loss tangent $tg(\delta)=0.032$ at \SI{100}{\giga \hertz} at \SI{4}{\kelvin}) have been taken from \cite{cr_100_lamb}. The input radiation is set as an \textit{Incident Gaussian Beam}, determined by the lens L1 feeding the blackbody cavity. The reflectance, computed as the ratio of the Poynting vector fluxes of the scattered and incident radiation, is $R(\SI{120}{\giga \hertz})=3.2\cdot 10^{-6}$. Diffractive effects and reflection coefficients at non-normal incidence, not included in the ray tracing approach, produce discrepancies between the two models. 

\begin{figure}[t]
\includegraphics[width=0.48\textwidth]{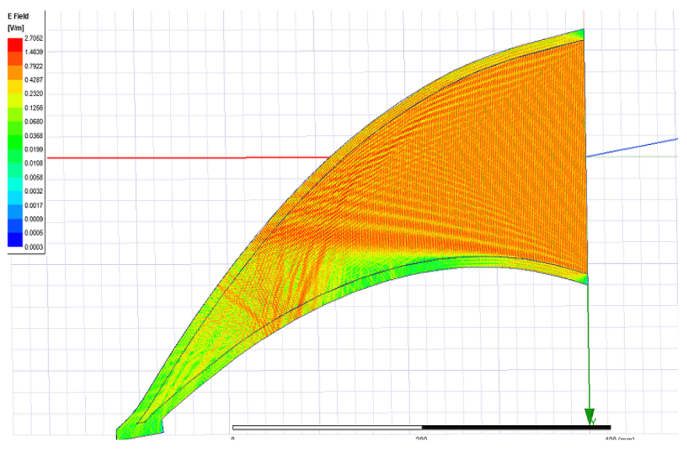}
\includegraphics[width=0.48\textwidth]{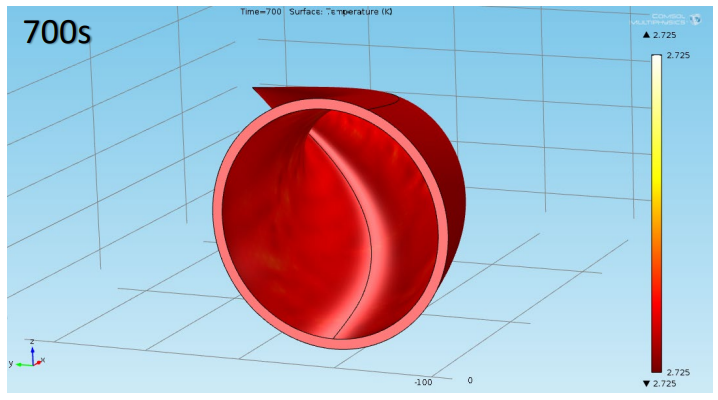}
\caption{{\bf Left:} Results of an HFSS simulation of the total electric field propagating within the blackbody cavity, demonstrating its gradual attenuation (colors correspond to a logarithmic scale for field amplitudes. {\bf Right:} Results of a finite elements thermal simulation displaying the temperature at the surface of the internal absorbing coating 700s after cool-down. The temperature gradients are $<$\SI{0.1}{\milli \kelvin}.}
\label{fig:hfss}
\end{figure}

Finite elements thermal simulations have been performed, considering a perfect thermal contact between the copper external structure and the internal absorbing coating, and including an irradiating internal surface of the cavity to a \SI{4}{\kelvin} background. The external structure temperature was set to \SI{2.725}{\kelvin}, from a starting point of \SI{4}{\kelvin}. The thermal properties of the coating are taken from \cite{LFI_cal}. Preliminary results show a maximum thermal gradient $<$\SI{0.1}{\milli \kelvin} (see fig.\ref{fig:hfss}).

\subsubsection{The atmospheric modulator}

The other input of the FTS looks at the sky through an optical window and a beam-steering mirror, which is used to modulate atmospheric emission and measure it as described in \S \ref{atmosphere}. The atmospheric scan has to be very fast (up to 2800 rpm) to measure atmospheric emission and subtract it in real time, so that 1/f noise from atmospheric emission is efficiently removed.  A light-weighted, all 6061 Aluminum design for the mirror has been optimized to minimize the inertia. The off-axis inertia moments are null using optimized counterweights, allowing us to use simple and reliable electrical motors and shaft bearings.

One of the most dangerous problems in this kind of measurements is the pickup of ground radiation. A custom forebaffle has been designed as the first defense against ground spillover. This forebaffle moves together with the receiver, and will be complemented by much larger steady ground shields. The internal surface will be fully covered with eccosorb absorber. The forebaffle is largely oversized, and has been shaped so that the spillover from its inner surface is not modulated by the spin of the wedge mirror. Heaters and temperature sensors will allow to test any residual spillover contribution to the measured signal by means of custom calibration measurements. 

\subsection{The cryogenic system}

A large cryogenic system, based on pulse-tubes and a $^3$He evaporation cooler is used to cool the optical system and the detector arrays. The system is similar to the one developed for the QUBIC experiment (see \cite{Masi21}), adapted for operation in the harsh Antarctic environment. To this purpose, we used the experience gained with the BRAIN refrigerator, successfully operated in Dome-C more than 10 years ago\cite{Polenta07}. In the right panel of figure \ref{fig:optics} we show the implementation of the COSMO instrument inside its cryogenic system. Using pulse tube refrigerators, the cryogenic system can operate continuously and does not rely on the delivery of cryogenic liquids, which would be very difficult in dome-C. Of course, continuous power is required, which amounts to about 10 kW, most of which are used for the pulse tube compressors and to warm-up the environment inside the experiment shelter to above 0 Celsius. 

\subsection{Readout electronics}\label{sec:electronics}

The arrays of KIDs are read applying a comb of frequencies tuned to the resonances of the different pixels, and reading the S21 parameter for each pixel. A FPGA-based electronic system generates a set of tones corresponding to the KIDs’ resonances. The combination of tones is sent to the detectors; the return signal is acquired and analyzed by means of a transceiver which detects the change in amplitude and phase of the tones. The system is based on a commercial architecture (Xilinx Virtex-5 NI7966R). The Virtex5 architecture was selected because it is the most powerful of the family being available as a rad-hard component. This specification not necessary for Antarctica, but is strategic in the perspective of the implementation of COSMO on a stratospheric balloon . We have prepared a state machine which, in the first state, performs a sweep to find the resonance tones of the KIDs. In the second state, a lookup table is generated with the values of the resonances found. With a CORDIC algorithm these tones are generated to compose a comb, which the third state of the machine transmits to the KIDs and acquires in amplitude and phase to measure the variation of the working point of the detectors. We have already tested the functionality of the state machine in the lab by means of Nb kids, fit for a 4 K cryostat. We are currently working to move the entire process of modulation and demodulation on board of the FPGAs, to reach the target rate of acquisition, fit for the operation with the FTS.

\section{Performance forecast for the the ground-based COSMO experiment}\label{sec:performance}

\begin{figure}[ht]
\centering
\includegraphics[width=0.97\textwidth]{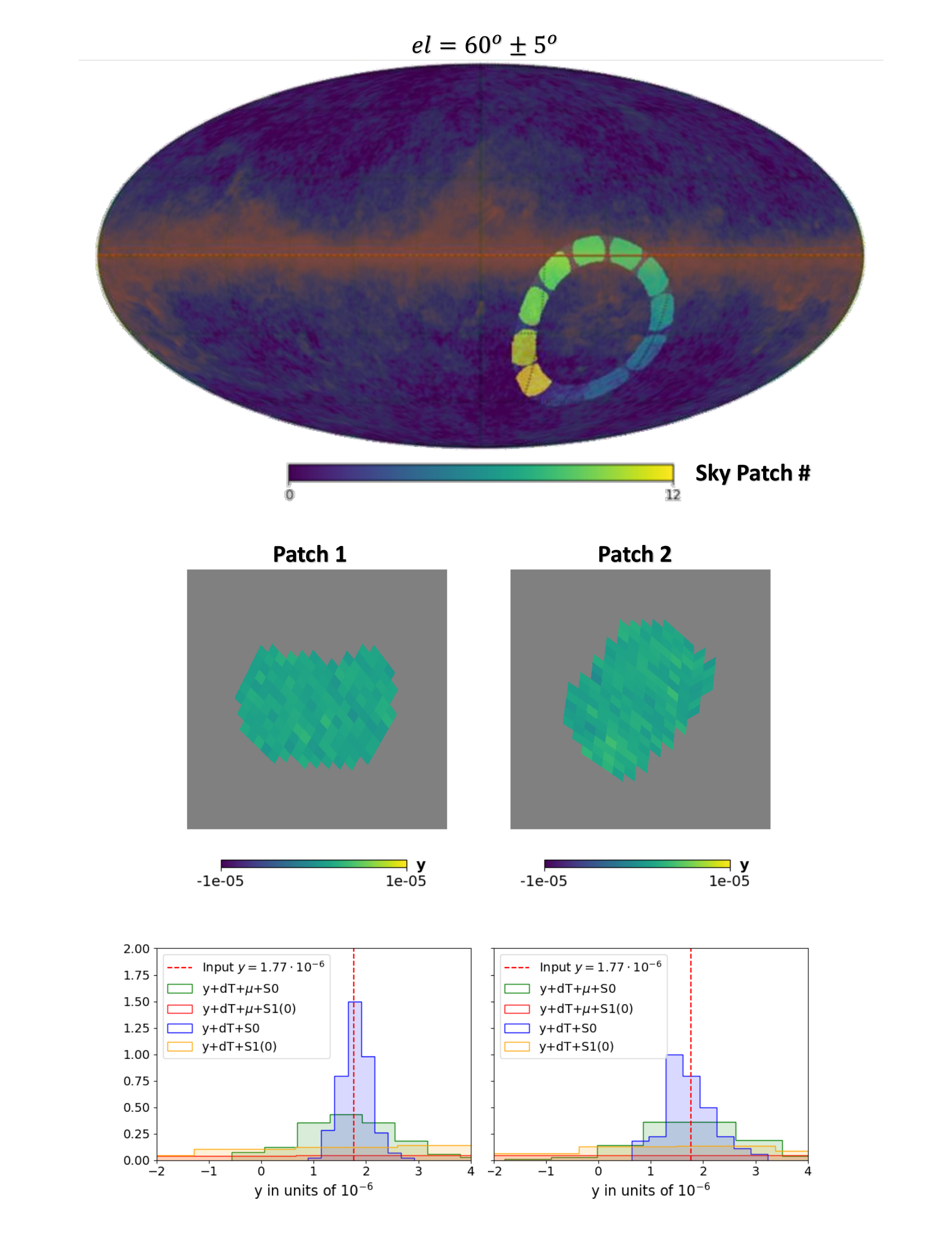}
\caption{{\bf Top:} Sky coverage map divided in 11 patches, overlapped to the 270GHz PySM map which includes thermal dust emission and CMB anisotropy. {\bf Middle:} Simulated measured maps for the $y$ distortion sky component, extracted using the ILC method, in two sample sky patches at high Galctic latitude. {\bf Bottom:} Histograms of the simulated estimates for the $y$ values, for the two sky patches above.}
\label{fig:ilc_sims}
\end{figure}

The performance of the instrument is assessed via ILC\cite{ilc} (Internal Linear Combination) based simulations. Absolutely-calibrated maps are used as input, containing the superposition of the PySM\cite{pysm} maps of CMB anisotropies and thermal dust, as the main galactic foreground. The isotropic distortion maps are added as a y-distortion map with $y=1.77\cdot 10^{-6}$ and a $\mu$-distortion map with $\mu=2.0 \cdot 10^{-8}$. 
The input multi-frequency maps, with a spectral resolution $\Delta \nu=\SI{15}{\giga \hertz}$, are smoothed with a FWHM=\SI{1}{\degree}. The noise realization, added as Gaussian noise, is dominated by the photon noise from the cryostat window, whose emission is modeled as a \SI{240}{\kelvin} grey-body with 1\% emissivity, and by the atmospheric emission (modeled based on the \textit{a.m.} software with PWV=0.15mm).

The ILC machinery allows to extract the y-distortion map from the dominant components. Different orders of the thermal dust emission are subtracted from the solution as in \cite{ilc_moment}. Histograms of the output Compton-y maps, reported in Fig.\ref{fig:ilc_sims}, show that the best ILC solution is provided by subtracting the $0^{th}$ order of the thermal dust, in combination with the CMB anisotropy. 

Assuming a scanning strategy at fixed central elevation, combined with the spinning wedge mirror modulation, $5\%$ of the sky is observed every day, which can be divided for the analysis in 11 sky-patches at different Galactic latitudes (see fig.\ref{fig:ilc_sims}). The ILC is independently applied to the patches and the weighted average of the output Compton-y maps provides the estimate of the isotropic y distortion as $y=(1.70 \pm 0.28)\cdot 10^{-6}$.  

\begin{figure}[t]
\centering
\includegraphics[width=0.48\textwidth]{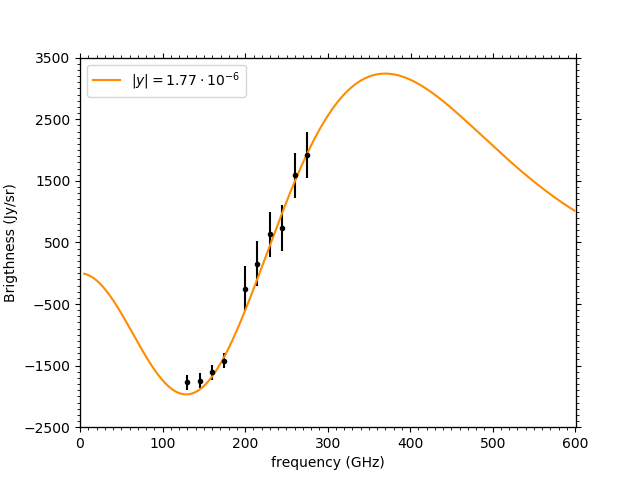}
\includegraphics[width=0.48\textwidth]{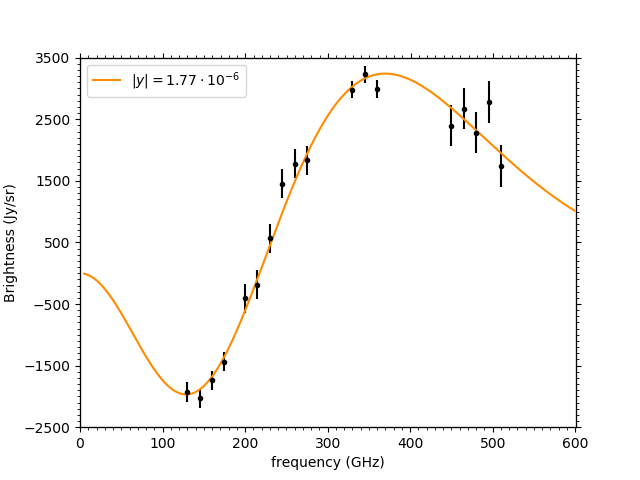}
\caption{{\bf Left:} Simulated measurement of the spectral distortion of the CMB after one year of integration in Dome-C, assuming photon noise limited performance dominated by the emissions of the cryostat window and the atmosphere and assuming \SI{15}{\giga \hertz} resolution. 
{\bf Right:} Simulated measurement of the spectral distortion of the CMB after 15 days of integration on a stratospheric balloon, assuming photon noise limited performance dominated by the CMB, residual atmosphere and warm mirror, and assuming \SI{15}{\giga \hertz} resolution.
}
\label{fig:spectrum-simulation}
\end{figure}

In figure \ref{fig:spectrum-simulation} (left panel) we show the expected spectrum of the spectral distortion we expect to measure after 1 year of integration in Dome-C. 

\section{Implementation of the balloon-borne COSMO experiment}\label{sec:balloonimplementation}

The spectrum of the CMB at mm wavelengths has been measured from stratospheric balloon platforms in the past (see e.g. \cite{Rich81}). Here we plan to exploit the recent advaces in detector sensitivity and speed, and the atmospheric modulation technique validated with the ground based implementation of COSMO, to take advantage of the very low residual atmospheric emission present at stratospheric altitude (see fig.\ref{fig:spectra}). With an atmospheric emission $\sim$300 times dimmer than in the ground based case, we expect that possible atmospheric residuals, which in the ground based implementation might escape the removal method  described in \S \ref{atmosphere}, will be negligible. In these conditions we expect that the limiting factor will become the emission of the warm components of the instrument, i.e. the room temperature sky scan mirror and the cryostat window. Due to the reduced atmospheric pressure at balloon altitude, the latter can be made very thin  (order of 25 $\mu$m), implying a factor $\sim$100 reduction of its thermal emission. We have a long-term plan to reuse the LSPE-SWIPE gondola and cryostat (\cite{Piac21}) to fly a modified COSMO instrument in the stratosphere in a time-scale of 4 years. This activity is synergic to the development of the BISOU experiment \cite{Maffei21}. Due to the reduced atmospheric load, we can extend our spectral coverage to significantly higher frequencies, in order to easy the separation of spectral distortions from local foreground emission. We rerun our simple simulation (\S \ref{sec:performance}) for the case of an observation in a 15-days flight, with photon-noise limited detectors (here the radiative background is dominated by the CMB, the thin window and the sky modulator mirror) obtaining the performance forecast shown in figure \ref{fig:spectrum-simulation} (right panel). The S/N ratio for the extraction of the largest spectral distortion is quite high ($>$10).

\section{Conclusions}\label{conclusions}

The measurement of spectral distortions in the CMB poses hard experimental challenges, but represents a fundamental cosmology tool, able to test a variety of cosmological phenomena. In particular, this kind of measurement seems to be the best, among very few, to shed light on the pre-recombination evolution of the universe. COSMO represents a pioneering sub-orbital approach to investigate spectral distortions at mm wavelengths. It copes with atmospheric emission and its fluctuations using fast detectors and a fast atmospheric emission modulation technique, able to separate its spectral brightness from the isotropic cosmological component. The instrument is in an advanced stage of development, and promises to detect the largest spectral distortion (the $y$ distortion due to ionized matter in the universe) in 2 years of integration from the French-Italian station of Concordia (Antarctica). A further implementation on a stratospheric balloon platform is under study, and promises significant advantages and a necessary validation step in view of the final space mission devoted to spectral distortions.

\section*{Acknowledgments}

The COSMO activity is funded in Italy by PNRA (National Program for Antarctic Research) and  PRIN (Programs with Relevant National Interest). The balloon-borne implementation is being studied within the COSMOS program of the Italian Space Agency. 

\bibliographystyle{ws-procs961x669}



\end{document}